\documentclass[a4paper,fleqn,usenatbib]{mnras}
\usepackage{amssymb,amsmath,graphicx,psfrag,mathtools}
\usepackage[T1]{fontenc} 
\usepackage{ae,aecompl} 
\usepackage{url}
\title[Are FRB Wandering Narrow Beams?]{Are FRB Wandering Narrow Beams?}
\author[J. I. Katz]{
J. I. Katz,$^{1}$\thanks{E-mail katz@wuphys.wustl.edu} 
\\
$^{1}$Department of Physics and McDonnell Center for the Space Sciences,
Washington University, St. Louis, Mo. 63130 USA 
}
\date{Accepted XXX.  Received YYY; in original form ZZZ} 
\pubyear{2016} 
\date{\today}
\begin{document} 
\label{firstpage} 
\pagerange{\pageref{firstpage}--\pageref{lastpage}} 
\maketitle 
\begin{abstract}
It is generally assumed that the sources of Fast Radio Bursts radiate
roughly isotropically, so that the observed low duty cycle of any individual
source indicates a similar low duty cycle of its radio-frequency emission.
An alternative hypothesis is that the radiative duty cycle is ${\cal O}(1)$,
but that the radiation is emitted in a beam with a solid angle comparable to
the observed duty cycle, whose direction wanders or sweeps across the
sky.  This hypothesis relaxes the extreme power demands of isotropically
radiating models of FRB at the price of multiplying the number of sources.
The constraints on pulsar models are relaxed; rather than being
unprecedentedly fast-spinning and highly-magnetized with short spin-down
times, their parameters may be closer to those of typical radio pulsars.
The nanoshots of Galactic pulsars pose analogous energetic problems that may
be resolved if their emission also is beamed, suggesting a relation between
these phenomena, despite their vastly different energy scales.
\end{abstract}
\section{Introduction}
\citet{L07} and \citet{T13} have estimated that fast radio bursts radiate as
much as $10^{40}$ ergs and set lower bounds on their radiated power as high
as $10^{43}$ ergs/s (because only upper bounds on their intrinsic durations
have been measured, their instantaneous power can only be bounded from
below).  These values are derived from their observed fluences and the
assumptions of cosmological distances and (at least roughly) isotropic
emission.  Distances are estimated by attributing FRB dispersions to
intergalactic plasma in standard cosmology.  The inferred brightness
temperatures are nearly as high as the highest astronomical brightness
temperatures ever measured, those of the nanoshots of two Galactic pulsars
\citep{S04,HE07}.

The evidence for cosmological distances is compelling \citep{K16a,K16b}, but
the assumption of isotropic emission has only simplicity and the lack of
contrary evidence to support it.  The high powers it implies pose demanding
constraints \citet{K16a} on models \citep{K12,CW15,CSP15,LBP16,K16b} that
attribute FRB to supergiant pulses from neutron stars with physical
processes like those of radio pulsars.  These models would otherwise be
attractive because they would naturally explain (assuming similar physical
processes, that unfortunately are not understood) the coherent emission with
extraordinary brightness temperatures of FRB.  These high brightness
temperatures also impose demanding constraints on models \citep{K14,RVV16}
of the emission mechanisms themselves.

In this paper I consider a hypothesis opposite to that of isotropic emission.
Suppose FRB radiate a steady power, but concentrate it into a beam of solid
angle $\Omega$ that sweeps or wanders in direction, and only intercepts the
observer a fraction $D$ of the time, where $D$ is the empirical duty factor.
$D$ may be roughly considered the fraction of the time the FRB is ``on'',
and is more formally defined as
\begin{equation}
D \equiv {\langle F(t) \rangle^2 \over \langle F(t)^2 \rangle},
\end{equation}
where $F(t)$ is the observed flux (or spectral density).  The duty factors
of several non-repeating FRB are roughly estimated $D \lesssim 10^{-8}$
\cite{K16b}, but the repeating FRB 121102 \citep{S16a,S16b} has $D$ as large
as $\sim 10^{-5}$.

If the emission is steady and if the direction of emission is randomly and
statistically isotropically distributed on the sky, the emission must be
collimated into a solid angle
\begin{equation}
\Omega \sim 4 \pi D\ \text{sterad}.
\end{equation}
The instantaneous power is less by a factor $D$ than it would be if
isotropic emission were assumed, but the brightness temperature, inferred
from the measured flux density, the distance and the source size, is the
same.  The requirement of coherent emission is not relaxed, but the
energetic constraints are mitigated.
\section{Energetic Constraints on Pulsar Parameters} 
\subsection{Constraints}
I follow the analysis of \citet{K16a}, but assume steady emission collimated
into the solid angle $\Omega$.  The most luminous FRB radiated a power
\citep{T13}
\begin{equation}
\label{power}
P \sim 10^{35} (E_{40}/\Delta t_{-3}) D_{-8}\ \text{erg/s}
\end{equation}
into a solid angle $\Omega \sim 4 \pi D$ in a pulse of equivalent isotropic
energy $10^{40} E_{40}$ ergs with a (not measured) unbroadened duration
$10^{-3} \Delta t_{-3}\ $s and duty factor $D \equiv 10^{-8} D_{-8}$.  $D$
is crudely bounded from limits on repetition of FRB other than the known
repeater FRB 121102 \citep{K16b}.  FRB 121102 has a larger $D \sim 10^{-5}$
but its outbursts were much fainter than the most intense bursts reported by
\citet{T13}, so the inferred mean $P$ is similar.

The power that can be emitted by a magnetic dipole rotating at an angular
frequency $\omega = 10^4 \omega_4$ s$^{-1}$, assuming the dipole axis is
perpendicular to the spin axis, is
\begin{equation}
\label{pdipole}
P_{dipole} = {2 \over 3} {\mu^2 \omega^4 \over c^3} \epsilon,
\end{equation}
where $\epsilon \le 1$ is an unknown efficiency of conversion of spin
energy into radio-frequency energy.  The magnetic dipole moment
\begin{equation}
\label{moment}
\mu \approx 10^{33} B_{15}\ \text{gauss-cm}^3,
\end{equation}
where $B = 10^{15} B_{15}$ gauss is the equatorial dipole field.

Combining Eqs.~\ref{power}--\ref{moment} yields
\begin{equation}
\label{dipoleparams}
B_{15}^2 \omega_4^4 \gtrsim 0.5 \times 10^{-15} {E_{40} D_{-8} \over \Delta
t_{-3} \epsilon} = 0.5 \times 10^{-15} {E_{40} \over \tau_5 \epsilon},
\end{equation}
where $\tau_5 \equiv \Delta t_{-3}/D_{-8}$ and $\tau = 10^5 \tau_5\,$s is
the mean interval between bursts.  $D$ is the duty factor describing the
actual burst duration at the source; the measured duration may be longer
because of instrumental response and propagation broadening.

\begin{figure}
\centering
\includegraphics[width=3in]{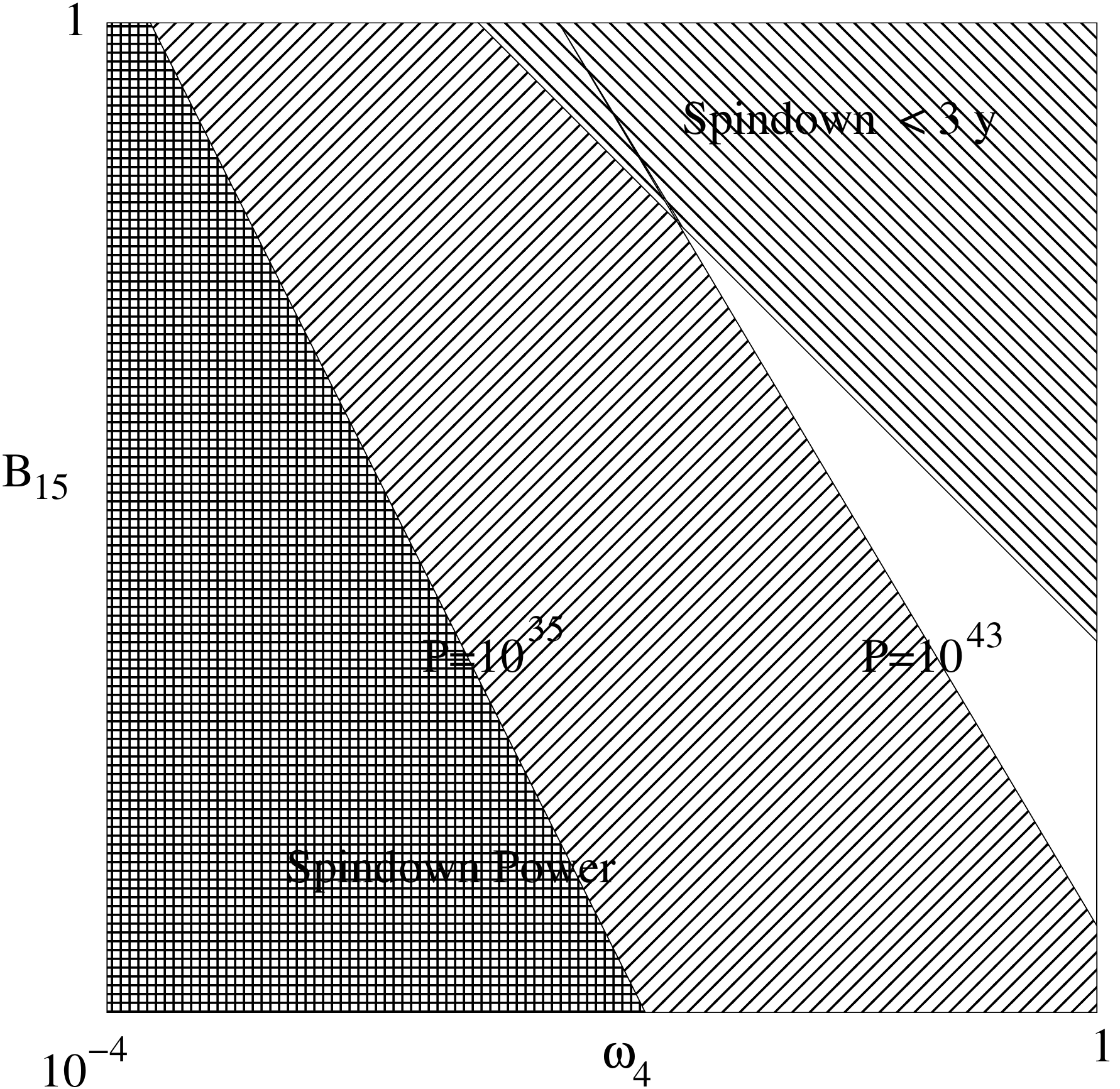}
\caption{\label{constraints} Constraints on magnetic field and spin rate in
pulsar models of FRB.  The region at the upper right is excluded by the
condition that the spindown time not be less than 3 y, the duration over
which the repetitions of FRB 121102 have been observed
\protect\citep{S16a,S16b}.  The heavily hatched region below and to the left
of the line $P = 10^{35}$ is excluded for narrowly beamed, randomly and
isotropically slewing, steady emitters with $E_{40}D/(\Delta t_{-3}\epsilon)
> 10^{-8}$, corresponding to the most energetic observed burst with duration
$\Delta t = 1\,$ms, efficiency $\epsilon = 1$ and duty factor $D = 10^{-8}$,
equivalent to a mean power of $10^{35}$ ergs/s.  The hatched region below
and to the left of the line $P = 10^{43}$ is excluded for isotropic emitters
with $E_{40}/(\Delta t_{-3}\epsilon) > 1$, corresponding to the most
energetic observed burst if it had duration $\Delta t = 1\,$ms and
efficiency $\epsilon = 1$.  Both factors in the denominator, $\Delta t_{-3}$
and $\epsilon$, are $\le 1$ because only upper limits to the intrinsic burst
durations are measured and the efficiency of turning rotational energy into
a radio burst must be $\le 1$.  If the bursts are unbeamed then the spindown
power $P \ge 10^{43}\,$ergs/s for the most luminous burst, confining the
possible parameters to the narrow unhatched region.  If they are strongly
beamed the wide swath between the $P = 10^{35}$ and $P = 10^{43}$ lines is
possible, provided the efficiency $\epsilon$ and duration in ms $\Delta
t_{-3}$ are not small.}
\end{figure}

Using the empirical lower bound of $T_{sd} \equiv 10^8 T_8\,\text{s} \gtrsim
10^8\,$s on the spindown time of FRB 121102 (recognizing that this may not
be applicable to other, more luminous, FRB) yields
\begin{equation}
\label{spindown}
B_{15}^2 \omega_4^2 \lesssim 10^{-5} T_8^{-1}.
\end{equation}
These constraints are illustrated in Fig.~\ref{constraints}.

Combining Eqs.~\ref{dipoleparams} and \ref{spindown} yields an upper bound
on the spin period, corresponding to the lower bound on $\omega_4$ of the
allowed region in Fig.~\ref{constraints}
\begin{equation}
\label{spin}
P_{spin} = {2 \pi \times 10^{-4}\,\text{s} \over \omega_4} \lesssim 100
\sqrt{\Delta t_{-3} \epsilon \over D_{-8} E_{40} T_8}\ \text{s} \approx 100
\sqrt{\tau_5 \epsilon \over E_{40} T_8}\ \text{s}.
\end{equation}
For beamed emission with $D = 10^{-8}$ periods as long as 100 s may bei
allowed, provided the magnetic field is large enough and depending on values
of the other parameters, but for isotropic emission $P_{spin} \lesssim
0.01\,$s.
\subsection{Comparison to known pulsars}
Even if FRB are produced by objects like radio pulsars (non-accreting
magnetic neutron stars), it is implausible that FRB parameters are in the
range of those of known Galactic radio pulsars that do not make FRB (or do
so too rarely to have been observed).  However, comparing known pulsar
parameters to the constraints on FRB pulsar parameters may give some insight
into what Nature may provide the modeler.

Catalogued pulsars \citep{ATNF} have values of $B_{15}^2 \omega_4^4$ in the
range $1.3 \times 10^{-22}$--$2 \times 10^{-12}$.  Many pulsars would
satisfy inequalities (\ref{dipoleparams}) and (\ref{spin}) if the product of
the dimensionless parameters were very roughly ${\cal O}(1)$, but all would
fail by several orders of magnitude for isotropic emission.

Values of $\epsilon$, tacitly assuming isotropic emission, in the range
$3 \times 10^{-2}$--$2 \times 10^{-11}$ are found in the ATNF Pulsar
Catalogue \citep{ATNF}, if the luminosity is evaluated at 400 MHz with an
assumed bandwidth of 400 MHz and is averaged over the pulse cycle.  The
value of $\epsilon$ corresponding to the peak of the pulse is likely to be
an order of magnitude higher.  However, these results cannot be directly
applied to the steady-source model of FRB because that model requires a
narrowly collimated beam.
\subsection{Nanoshots}
The peak nanoshot luminosities of PSR B1937+21 \citep{S04} and of the Crab
pulsar \citep{HE07}, assuming isotropic emission, roughly equal their
spindown powers.  This would imply $\epsilon \approx 1$, but the assumption
of isotropic emission would make that value of $\epsilon$ inapplicable to
steady-source (beamed) models of FRB.  

\citet{S04}, assuming isotropic emission, inferred a radiation energy
density in the nanoshots of PSR B1937+21 that exceeded (if they were emitted
from a region of dimension $\sim (c \Delta t) \sim 500\,$cm) the plasma
energy density by a factor of $\sim 300$, and that was even greater than the
magnetostatic energy density.  This is implausible: a pulsar can radiate its
rotational energy only at the dipole radiation rate and the radiation
Poynting vector is distributed over the open field lines.  They cover a
fraction ${\cal O}(\omega R/c) \sim {\cal O}(0.1) \sim 10^{12}\,$cm$^2$ of
the surface of a fast pulsar ($P = 1.557\,$ms) like B1937+21, so that only
$\sim (500\,\text{cm})^2/10^{12}\,\text{cm}^2 \sim 2.5 \times 10^{-7}$ of
the spindown power flows through the apparent $\sim 500\,$cm source region
($\Delta t \approx 15\,$ns) of the nanoshot.  The spindown power flowing
through the apparent source region is several orders of magnitude less than
the inferred (isotropic) radiated power.

It is implausible that a nanoshot could be powered by the neutron star's
magnetic energy because this can only be tapped by magnetic reconnection.
This occurs in young high field neutron stars (``magnetars'') and is
observed as Soft Gamma Repeaters and Anomalous X-ray Pulsars.  These emit
thermal X-rays at temperatures of tens of keV, unlike the coherent radio
emission of nanoshots or FRB.

A plausible resolution of this energy dilemma is that the nanoshots are
highly collimated, reducing the radiation energy density, perhaps by many
orders of magnitude, compared to that implied by isotropic emission.  They
may be Galactic exemplars of the highly collimated pulsar emission that this
paper hypothesizes for FRB, though FRB are $\sim 10^{13}$ times as energetic
as the nanoshots (that have $\sim 0.1$ of the fluence of FRB and are $\sim
10^{-6}$ times as distant).  Despite this enormous difference of energy
scale, the brightness temperatures of nanoshots and FRB are within about two
orders of magnitude of each other, suggesting a similarity.
\section{Collimated Radiation}
In Nature, collimated radiation is only produced by collimated beams of
relativistic particles.  Radiation is generally broadly distributed in angle
(for example, in a dipole radiation pattern) in the instantaneous rest
frames of the emitting particles, and Lorentz transformation to the
observer's frame collimates it to an angle $\theta \sim 1/\Gamma$, where
$\Gamma$ is the emitters' Lorentz factor.  In the laboratory collimated
radiation is produced by engineered structures (laser cavity mirrors,
lenses, absorbing collimators {\it etc.\/}), but these are not likely to
occur naturally.  Intensity spikes may occur at caustics, but in propagation
through turbulent media many individual ray paths add randomly and the
intensity has a Rayleigh distribution.  Intensities orders of magnitude
greater than the mean are exponentially rare.

Collimation to a fraction $D$ of the sphere indicates emission by
relativistic particles with Lorentz factors $\Gamma \sim 1/(4 \pi D)$ if the
particles (and radiation) are confined to a thin but broad sheet (such as
a surface of constant magnetic latitude; such structures are observed in
auror\ae) and Lorentz factors $\Gamma \sim 1/\sqrt{4 \pi D}$ if the
particles and radiation form a pencil beam.  The observed duty factors of
FRB suggest $\Gamma \sim 10^5$--$10^8$ for sheeds and $\Gamma \sim
300$--$10^4$ for pencil beams.
\section{Detecting Periodicity}
If repeating FRB can be shown to be periodically modulated their production
by rotating neutron stars will be demonstrated.  The values of the period
and its rate of change will distinguish pulsar from SGR models and determine
their parameters.  Phasing the bursts of repeating FRB may detect
periodicity {\it provided\/} their emission is modulated by their rotation
as well as by the wandering that produces rare and sporadic bursts when the
assumed narrow beam points in our direction.  Such periodic modulation is
certainly plausible (it is the characteristic signature of classic radio
pulsars), but by no means certain.  It might manifest itself in a
periodogram of the epochs of observed bursts.

Suppose we observe $N$ bursts from a FRB in a time $T_{obs}$ and wish to
search for modulation frequencies over a bandwidth $\nu$ (typically from 0
to $\nu$).  If all pulses are equally strong (deviations from this will
reduce the effective value of $N$) and all observable pulses occur at the
same rotational phase of the pulsar then the effective number of independent
samples (independent elements of a periodogram) is about $2\nu T_{obs}$.  If
the observed pulse times are distributed as shot noise (the null hypothesis
of no periodic modulation) then the elements of the periodogram have a
Gaussian distribution (if $N \gg 1$) with mean $\mu = 0$ and standard
deviation $\sigma = \sqrt{N}$.  If the pulses are exactly periodic the value
of the element of the periodogram corresponding to that period is $N$.  In
order to have a probability $p \ll 1$ of a false positive period, $N$ must
satisfy
\begin{equation}
N > \ln{\left({2 \nu T_{obs} \over p}\right)}.
\end{equation}
For an observing run with $T_{obs} = 2 \times 10^4\,$s (about 6 hours),
$\nu = 1000/$s and $p = 0.003$ ($3\sigma$ detection) $N > 24$ is required.
For $p = 3 \times 10^{-7}$ ($5\sigma$ detection) $N > 32$ is required.
These values are about an order of magnitude greater than those detected
by \citet{S16b} and \citet{S16a} in their studies of the repeating FRB
121102, but small enough that it is plausible that sufficiently many pulses
will be observed in the future, perhaps by FAST or SKA, more sensitive than
any existing instruments.

These estimates assume a constant pulse period, a good approximation for
$T_{obs} \ll 1\,$day, even for rapidly slowing pulsars.  However, for
data scattered over many days it may be necessary to allow for spin-down
$\dot \nu$, and perhaps even higher derivatives; the phase
\begin{equation}
\label{phase}
\phi = \phi_0 + \nu t + {1 \over 2} {\dot \nu} t^2 + {1 \over 6} {\ddot \nu}
t^3 + \ldots.
\end{equation}
This increases the dimensionality of the parameter space to be searched and
the required number of detected pulses increases rapidly.  Timing noise can
make it impossible to detect underlying periodicity because then the series
\ref{phase} does not converge.  This may occur when the detected signal
consists, as for FRB and RRAT, of rare but accurately timed pulses rather
than, as for most radio pulsars, a continuously modulated periodic signal.
\section{Discussion}
The assumption that FRB emit roughly isotropically with a very low duty
factor represents one end of a continuum.  In rotation-powered pulsars there
is no energy reservoir intermediate between the neutron star's rotation and
the radiation field, so these assumptions combine to impose extreme
constraints \citep{K16a} on pulsar parameters.  This paper has explored the
opposite limit, in which FRB emit steadily in a narrow beam that only rarely
points to the observer.  It is not possible to distinguish these limits
phenomenologically, but the assumption of steady collimated emission relaxes
the constraints on the pulsar parameters, and is therefore attractive
theoretically.  The truth may lie somewhere between these extremes.

If there were a complete model of coherent emission in FRB it would be
required to explain their high brightness.  In the absence of such a model I
abdicate this responsibility, and only consider the easier problem of
energetic constraints.  This might be excused by comparison with classic
radio pulsars:  The origin of their energy in spindown of rotating neutron
stars is understood, but their coherent emission is not, despite a
half-century of effort.

\bsp 
\label{lastpage} 
\end{document}